\documentclass[conference]{IEEEtran}
\IEEEoverridecommandlockouts

\usepackage{cite}
\usepackage{tikz}
\usepackage{todonotes}
\usetikzlibrary{arrows.meta, positioning}
\usepackage{url}
\usepackage{hyperref}
\usepackage{booktabs}
\usepackage{pgf}
\usepackage{amsmath}
\usepackage{amsfonts}
\usepackage{graphicx}
\usepackage{float}
\usepackage{xcolor}
\usetikzlibrary{
    shapes,
    arrows,
    positioning,
    calc,
    fit,
    backgrounds,
    decorations.pathreplacing,
    decorations.markings,
    patterns,
    shadows,
    matrix
}

\usepackage{caption}
\definecolor{serviceblue}{RGB}{225, 245, 254}
\definecolor{storageviolet}{RGB}{243, 229, 245}
\definecolor{externalorange}{RGB}{255, 243, 224}
\definecolor{processgreen}{RGB}{232, 245, 232}

\tikzstyle{service} = [rectangle, rounded corners, minimum width=3cm, minimum height=1cm, text centered, draw=black, fill=serviceblue, drop shadow]
\tikzstyle{storage} = [cylinder, shape border rotate=90, minimum width=2.5cm, minimum height=1.5cm, text centered, draw=black, fill=storageviolet, drop shadow]
\tikzstyle{external} = [ellipse, minimum width=2.5cm, minimum height=1cm, text centered, draw=black, fill=externalorange, drop shadow]
\tikzstyle{process} = [rectangle, minimum width=2.5cm, minimum height=1cm, text centered, draw=black, fill=processgreen, drop shadow]
\tikzstyle{arrow} = [thick,->,>=stealth]
\tikzstyle{bidirectional} = [thick,<->,>=stealth]
\title{\emph{Mark My Works} \\ Autograder for Programming Courses}

\author{\IEEEauthorblockN{Yiding Qiu, Seyed Mahdi Azimi, Artem Lensky}\\
  \IEEEauthorblockA{School of Engineering and Technology, UNSW Canberra, Australia\\
  \{yiding.qiu, s.azimi, a.lenskiy\}@unsw.edu.au}
  }

\begin{document}
\maketitle

% -----------------------------------------------------------------------------
\begin{abstract}
Large programming courses struggle to provide timely, detailed feedback on student code. We developed \emph{Mark My Works}, a local autograding system that combines traditional unit testing with LLM-generated explanations. The system uses role-based prompts to analyze submissions, critique code quality, and generate pedagogical feedback while maintaining transparency in its reasoning process. 

We piloted the system in a 191-student engineering course, comparing AI-generated assessments with human grading on 79 submissions. While AI scores showed no linear correlation with human scores ($r$ = -0.177, $p$ = 0.124), both systems exhibited similar left-skewed distributions, suggesting they recognize comparable quality hierarchies despite different scoring philosophies. 
The AI system demonstrated more conservative scoring (mean: 59.95 vs 80.53 human) but generated significantly more detailed technical feedback.
\end{abstract}


\section{Introduction}
The increase in enrollment in programming courses places considerable stress on assessment pipelines.  In Computational Problem Solving (ZEIT1307) at UNSW Canberra, which is a first-year engineering course, students submit totally almost 600 Jupyter notebooks per semester that require marking.  First, marking 600 notebooks take considerable amount of time. Second, different tutors apply the rubric unevenly. Third, students now expect feedback well before the deadline. A typical solution for large programming courses is unit-tests. Although, unit-tests do address factual correctness, they rarely explain why a solution fails, nor do they comment on style, documentation, or design. Commercial platforms such as Gradescope and CodeRunner embed unit tests and plagiarism checks, yet remain dependent on cloud services and offer limited narrative commentary.  At the other extreme, emerging research prototypes employ LLMs to critique code but raise data‑governance concerns and may hallucinate.  

Mark My Works offers a privacy-scalable deployment model: run the stack on campus or in an institution-owned AWS account for elastic compute—no vendor lock-in. A rubric-aware prompt chain keeps every score traceable: the model judges each atomic rubric item, critiques code structure, and rewrites its reasoning as plain-language advice. This three-role workflow yields (1) objective marks tied to explicit criteria, (2) consistent grading across large cohorts, and (3) rapid, actionable feedback delivered minutes after submission.

% -----------------------------------------------------------------------------
\section{Related Work}
There are several marking systems available online. For example, Gradescope scales well via Docker tests, but its feedback stops at a pass/fail log\cite{gradescope}, whereas CodeRunner integrates unit tests into Moodle quizzes for immediate correctness feedback\cite{coderunner}. PrairieLearn emphasises parameterised question generation and mastery learning, yet its feedback is mostly pass/fail test logs\cite{prairielearn}. Autolab offers an open‑source, on‑premises autograder with scoreboard visualisations yet minimal natural‑language commentary\cite{autolab}. Vocareum delivers scalable cloud Jupyter environments and peer or TA grading but depends on external hosting\cite{vocareum}. GitHub Classroom enables repository‑based autograding scripts inside CI pipelines, simplifying workflow integration but limiting real‑time narrative guidance\cite{githubclassroom}. Commercial CodeGrade focuses on inline code annotations and LMS integration, though detailed pedagogical feedback is largely tutor-authored\cite{codegrade}.

Recent studies explore \textit{LLM evaluation of code quality and style} \cite{pankiewicz2023gptfeedback} and chain‑of‑thought
reasoners to explain errors\cite{finnie2022robots}. However, LLM are known to suffer from hallucination\cite{huang2023survey}, underscoring the need
for deterministic test scaffolds and human‑in‑the‑loop review.

Compared with these efforts, \href{https://MarkMyWorks.com}{MarkMyWorks.com} integrates explicit rubric‑role prompts and targets open‑ended engineering programming tasks whose correct outputs are continuous‑valued-such as floating‑point numbers, arrays, or plots-rather than discrete multiple‑choice selections.

\section{Methods}
\textit{Experimental setup.} The system was piloted during the 2025 academic year in Computational Problem Solving (ZEIT1307), a first-year programming course at UNSW Canberra, with 191 students participating. The cohort comprises Australian Defence Force Academy (ADFA) Defence Trainee Officers and Cadets alongside civilian engineering students. Their programming experience ranges from zero to year-12 software studies. Students submit solutions as Jupyter notebooks containing both executable code and explanatory markdown, requiring assessment of algorithmic correctness, mathematical reasoning, and communication skills.

Two representative assignments were selected for evaluation: Assignment 2 consists of two parts, focusing on numerical integration and data analysis. Part 1 requires students to implement area calculation algorithms from first principles and part 2 asks to develop regression analysis using matrix operations. Assignment 3 addresses non-linear model fitting and optimization, where students choose between technology adoption modeling using SciPy's \texttt{curve\_fit} or implementing gradient descent algorithms for epidemic wave modeling. Both assignments emphasize modular code design, mathematical understanding, and interpretation of results within engineering contexts.

\textit{Pilot evaluation constraints.} Due to computational resource limitations and pipeline development timeline constraints, AI assessment was applied to 120 student submissions during the initial deployment phase. 
From this subset, 79 submissions successfully completed the full pipeline evaluation, while 41 submissions were flagged for manual review and assigned preliminary scores of zero by the system's quality assurance protocols.

% The marking system implements comprehensive submission integrity checks designed to ensure fair evaluation. Submissions are automatically flagged for human intervention when the pipeline detects file corruption, extraction failures, incomplete content, prompt injection attempts, unknown behaviors, or submissions containing custom environments that could compromise assessment reliability. This conservative approach prioritizes evaluation fairness over coverage, ensuring that only submissions meeting technical requirements receive automated scoring.

% The 79 successfully processed submissions represent a diverse cross-section of student performance levels and programming approaches, ranging from basic implementations to sophisticated solutions. This subset excludes submissions with technical issues that could artificially inflate or deflate scores, providing a clean dataset for evaluating the AI system's assessment capabilities under normal operating conditions.

% \todo[inline]{Marking rubric}% yiding
\textit{Marking rubric.} Traditional monolithic rubrics present challenges for automated processing due to subjective criteria and interdependent assessment dimensions. \href{https://MarkMyWorks.com}{\textit{Mark My Works}} addresses this through \emph{atomized assessment modules}, where complex assignments are decomposed into discrete, evaluable components that map directly to system processing units.

Each assignment is configured through a structured YAML specification that defines modular assessment workflow. For Assignment 3 (non-linear modeling), the 100-point rubric decomposes into 24 independent modules across six functional categories: data preparation (10 points), model implementation (35 points), optimization algorithms (25 points), results analysis (15 points), code quality (15 points). Each module encapsulates specific assessment criteria with structured input/output interfaces, enabling parallel processing and consistent scoring.

The modular architecture allows targeted LLM prompts tailored to specific competencies. For instance, the \texttt{p1\_parameter\_interpretation} module receives fitted model parameters and student explanations, focusing exclusively on conceptual understanding of mathematical relationships. In contrast, \texttt{p2\_optimization\_implementation} evaluates algorithmic correctness by analyzing code structure and constraint handling. This decomposition transforms subjective "\textit{understanding}" into concrete, assessable artifacts while maintaining pedagogical coherence through the final report assembly module.

\section{Mark My Works design}
\subsection{Marking system}
% \todo[inline]{System diagram and description of each module and the problems that each module address}% yiding - mechanisms other than llm
% \begin{figure}[H]
% \centering
% \includegraphics[width=\textwidth]{diagrams/system_overview_mermaid.pdf}
% \caption{System Overview Architecture showing the modular service design}
% \label{fig:system_overview}
% \end{figure}

\begin{figure}[]
\centering
\includegraphics[width=0.44\textwidth]{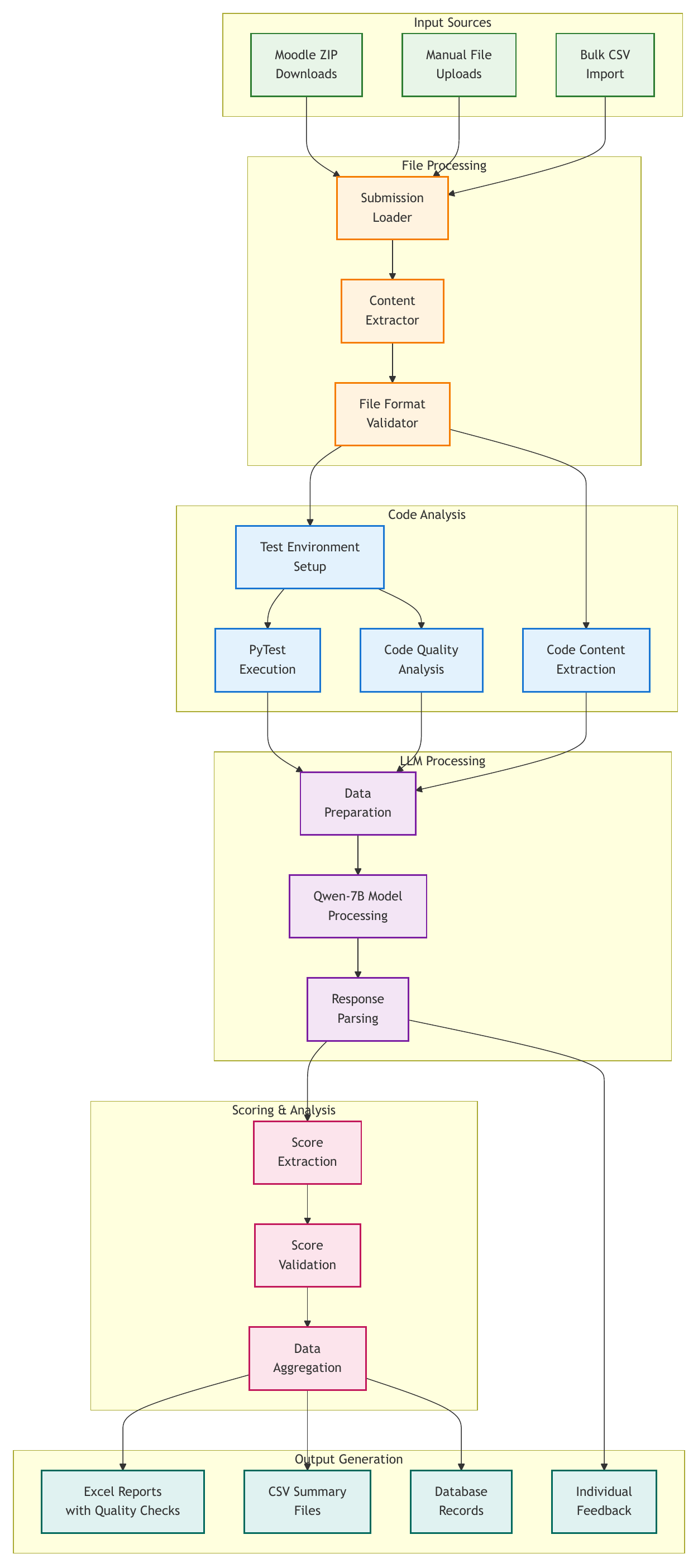}
\caption{Data Flow Pipeline from submission to final reports}
\label{fig:data_flow}
\end{figure}

The Mark My Works system addresses the three core challenges of scale, consistency, and formative demand through a modular pipeline architecture. Figure~\ref{fig:data_flow} illustrates the end-to-end processing flow from submission intake to comprehensive feedback generation.

The system processes submissions through six core modules: Input Sources handle multi-modal submission intake (Moodle, web upload, CSV), File Processing standardizes heterogeneous formats, Code Analysis applies uniform testing and quality metrics, LLM Processing generates explanatory feedback through role-based prompts, Scoring validates and extracts grades, and Output Generation produces stakeholder-specific reports \cite{palahan2025pythonpal}.

Each module operates independently, allowing for parallel processing and fault isolation. When individual student submissions fail (e.g., due to syntax errors or missing files), the pipeline continues processing other submissions, generating partial feedback where possible and flagging problematic cases for manual review. This design maintains system reliability while preserving the pedagogical goal of providing feedback to all students, even those with incomplete or malformed submissions.

% \begin{figure}[H]
% \centering
% \includegraphics[width=\textwidth/2]{diagrams/marking_pipeline_mermaid.pdf}
% \caption{Marking Pipeline Internal Architecture and component interactions}
% \label{fig:marking_pipeline}
% \end{figure}
% env diff
\subsection{User Interface and Backend}
The webUI of \href{https://MarkMyWorks.com}{\textit{Mark My Works}} platform provides an interface for both students and tutors, developed using React. The backend is implemented in Python using the FastAPI framework.

Student submissions are manually uploaded through the website and the integration module with external learning management systems like Moodle is being developed. Submitted files are stored in a local MinIO object storage system, while metadata for users, courses, tasks, submissions, and feedback is persisted in a PostgreSQL database. Also, the autograding tasks are queued using Celery with Redis as the message broker to ensure non-blocking execution.

\begin{figure}[H]
\centering
\caption{System Architecture showing the modular service design}
\includegraphics[width=0.5\textwidth]{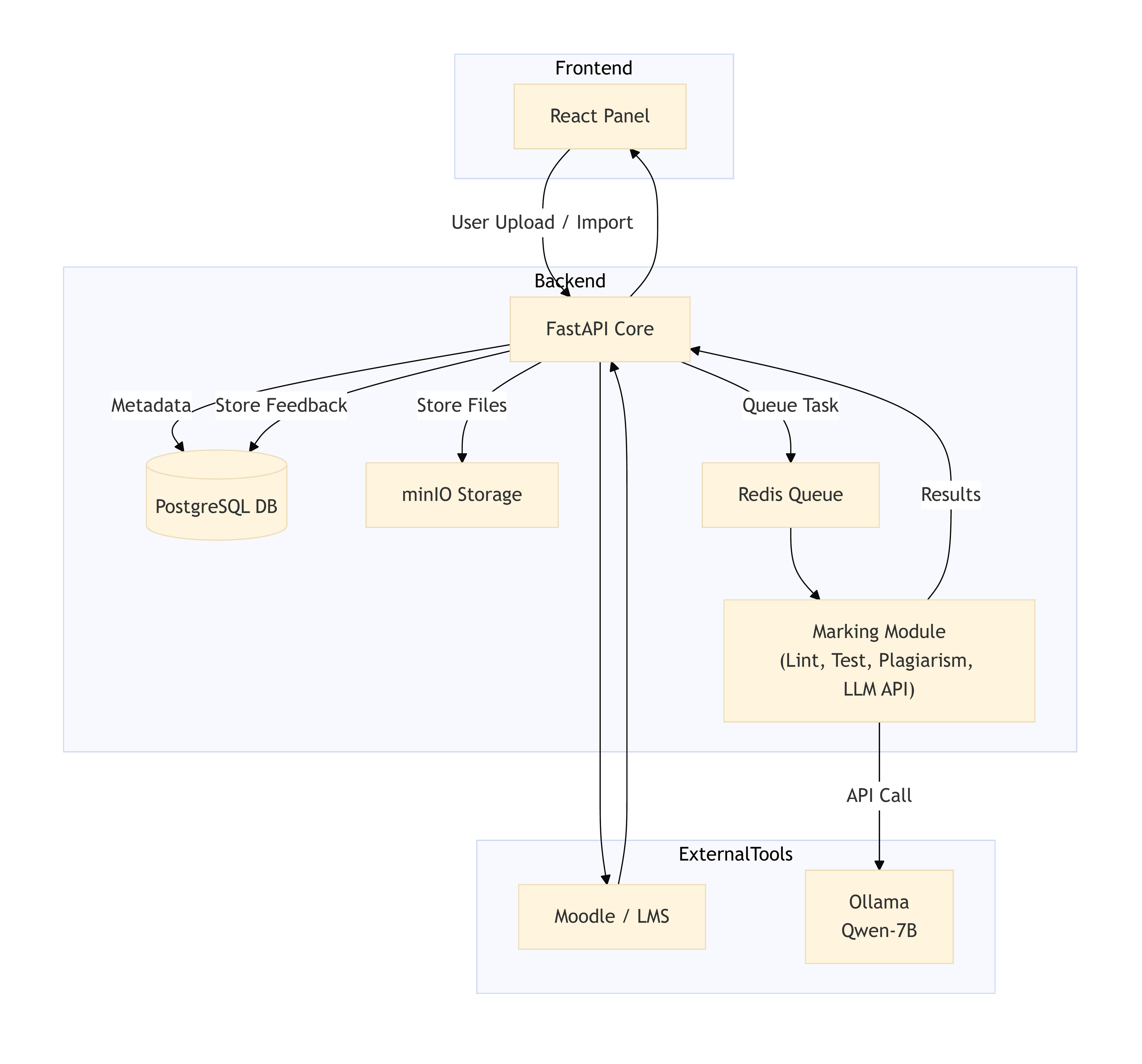}
\label{fig:software_arch}
\end{figure}

The queued jobs are handled by a Marking Module as described above performs automated grading and generates a report. The resulting scores and structured feedback are saved in the database and displayed to the corresponding users via the React frontend. This decoupled architecture supports flexible deployment, improves maintainability, and enables deeper future integrations with other assessment platforms or LLM models.
Figure \ref{fig:software_arch} illustrates the overall system architecture and the flow of data between components.

\section{Results}
We evaluated the Mark My Works system by comparing AI-generated scores with human assessments across two assignments in ZEIT1307. 

\subsection{Sample Methodology}
The evaluation was conducted on Assignment 2 and Assignment 3 submissions from the 2025 offering of ZEIT1307. Human grading was performed by trained tutors following the rubrics, resulting in 184 scored submissions. AI assessment was subsequently applied to a subset of 79 submissions, selected based on submission completeness and format compatibility with the automated processing pipeline. This subset represents submissions that contained all required components (code cells, markdown explanations, and data files) necessary for comprehensive AI evaluation. Zero scores, indicating incomplete or missing evaluations, were excluded from both datasets to ensure meaningful comparison.

\subsection{Raw Score Statistics}
Table~\ref{tab:raw_statistics} presents the descriptive statistics for both human and AI assessments before any normalization. The human scores exhibit a higher mean (80.53) with lower variance ($\sigma$ = 12.88), while AI scores show a lower mean (59.95) with higher variance ($\sigma$ = 17.94). Both distributions demonstrate left-skewed characteristics (human: -1.724, AI: -1.409), indicating concentration in higher score ranges despite the different absolute values.

\begin{table}
\centering
\caption{Raw Score Statistics: Human vs. AI Assessments}
\begin{tabular}{lcc}
\hline
\textbf{Metric} & \textbf{Human ($n$=184)} & \textbf{AI ($n$=79)} \\ \hline
Mean           & 80.53                   & 59.95              \\
Median         & 83.00                   & 66.41              \\
Std Deviation  & 12.88                   & 17.94              \\
Skewness       & -1.724                  & -1.409             \\
Min Score      & 42.00                   & 15.63              \\
Max Score      & 100.00                  & 87.50              \\ \hline
\end{tabular}
\label{tab:raw_statistics}
\end{table}

\subsection{Correlation Analysis}
\begin{figure}[H]
\centering
\caption{Human Score vs AI Score with Regression Analysis \\ Correlation ($r$): -0.177, $p$-value: 0.124 \\ Regression: Y = -0.25X + 79.71}
\includegraphics[width=\textwidth/2]{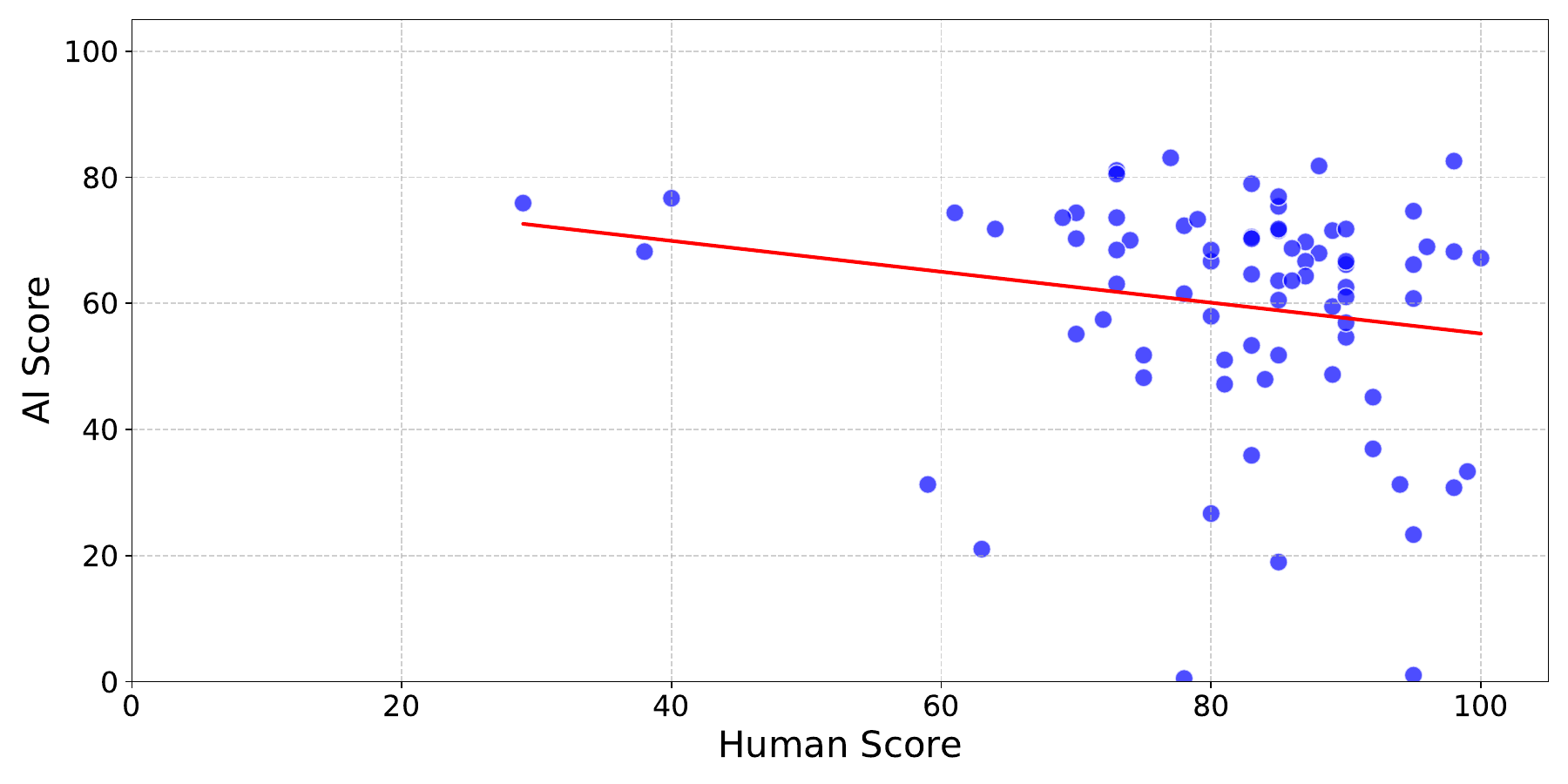}
\label{fig:human_vs_ai_regression}
\end{figure}

To assess the relationship between human and AI scoring patterns, we calculated the Pearson correlation coefficient using raw scores from the overlapping subset of 79 submissions. Figure~\ref{fig:human_vs_ai_regression} presents the scatter plot with fitted regression line.

The analysis reveals a weak negative correlation ($r$ = -0.177) that is not statistically significant ($p$ = 0.124). This suggests no meaningful linear relationship between human and AI scoring patterns. The negative trend suggests that submissions receiving higher human scores tend to receive slightly lower AI scores, though this relationship is not reliable for predictive purposes.

\subsection{Normalization and Scale Alignment}
Given the different scoring ranges observed in raw data (human: 42-100, AI: 15.63-87.50), we applied min-max normalization to enable distributional comparison. The transformation extends AI scores to match the human scoring scale:

\begin{equation}
\texttt{Norm}_{AI} = \frac{S_{AI} - \min(S_{AI})}{\max(S_{AI}) - \min(S_{AI})} \cdot \max(S_{human})
\end{equation}

This normalization preserves the relative ordering and distributional shape of AI scores while enabling visual comparison with human score distributions. Table~\ref{tab:normalized_statistics} shows the effect of this transformation.

\begin{table}[h!]
\centering
\caption{Comparison of Raw and Normalized AI Scores}
\begin{tabular}{lcc}
\hline
\textbf{Metric} & \textbf{AI Raw} & \textbf{AI Normalized} \\ \hline
Mean           & 59.95           & 71.73                  \\
Std Deviation  & 17.94           & 22.09                  \\
Min Score      & 15.63           & 0.00                   \\
Max Score      & 87.50           & 100.00                 \\ \hline
\end{tabular}
\label{tab:normalized_statistics}
\end{table}

\subsection{Distributional Shape Comparison}
\begin{figure}[H]
\centering
\caption{Distribution of Human, Raw AI, and Normalized AI Scores}
\includegraphics[width=\textwidth/2]{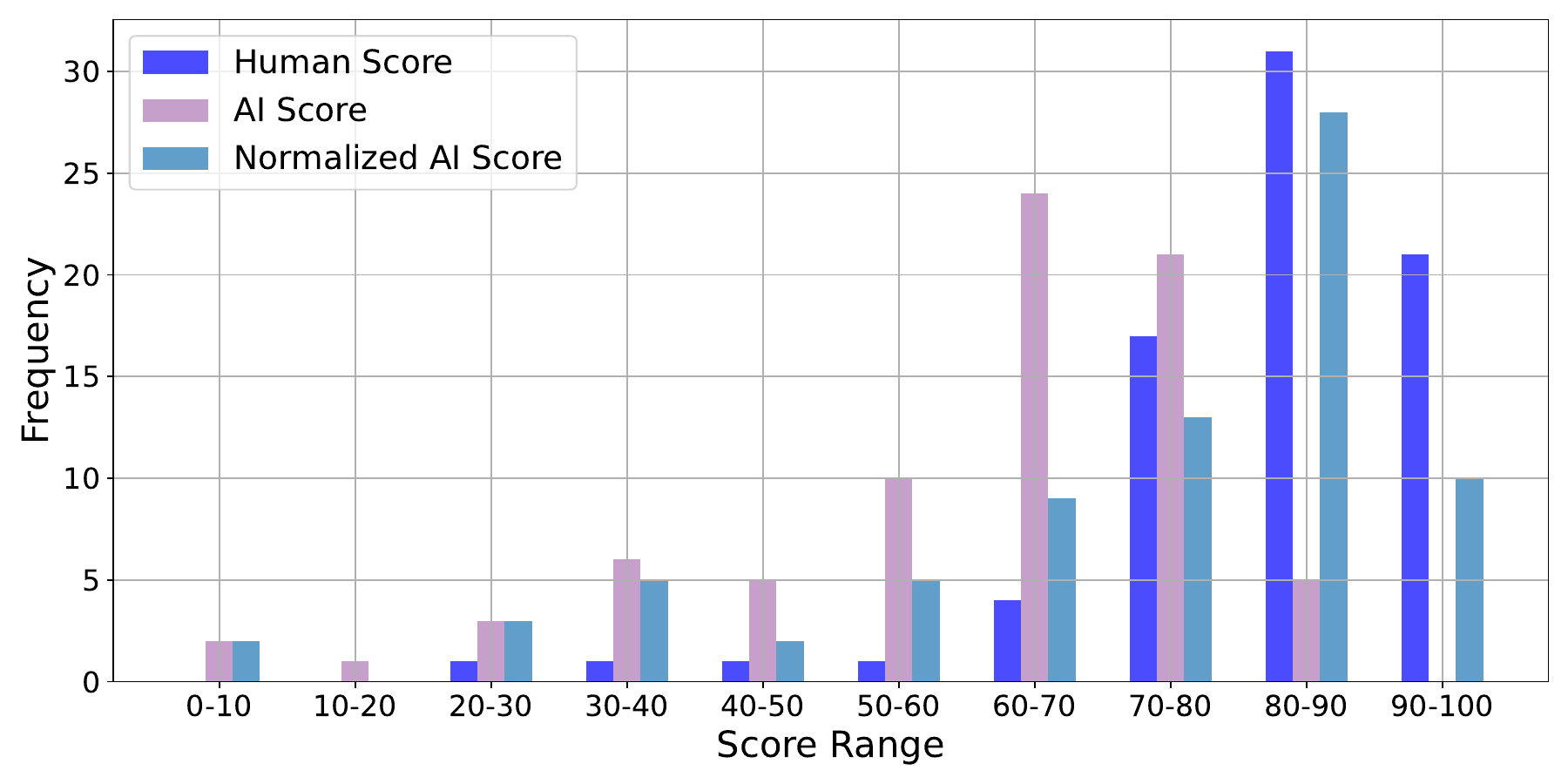}
\label{fig:score_distribution}
\end{figure}

Figure~\ref{fig:score_distribution} illustrates the score distributions across 10-point intervals for human scores, raw AI scores, and normalized AI scores. Despite the lack of linear correlation, both human and AI distributions exhibit similar left-skewed shapes, with peaks in higher score ranges and comparable decline patterns.

The distributional similarity suggests that while human and AI systems may weight different aspects of code quality, both recognize a fundamental distinction between higher and lower quality submissions. Human scores concentrate heavily in the 80-100 range, while AI scores (both raw and normalized) show broader distribution with fewer submissions in the highest brackets, indicating more conservative scoring behavior.

\section{Discussion}

The evaluation reveals a complex relationship between human and AI assessment approaches that merits careful interpretation. While the systems demonstrate fundamentally different scoring behaviors, several findings suggest potential complementary value in programming education assessment.

\subsection{Assessment Approach Differences}
The weak negative correlation ($r$ = -0.177, $p$ = 0.124) between human and AI scores indicates that the two assessment methods evaluate different aspects of student submissions or weight assessment criteria differently. This lack of linear relationship should not be interpreted as system failure, but rather as evidence of distinct evaluation philosophies. Human evaluators may prioritize conceptual understanding, partial credit for effort, and pedagogical considerations, while the AI system applies rubric criteria more systematically without contextual adjustments.

The conservative AI scoring pattern (mean: 59.95 vs. 80.53 human) combined with higher variance ($\sigma$ = 17.94 vs. 12.88) suggests the AI system applies assessment criteria more stringently and consistently across submissions. This behavior may reflect the absence of \emph{leniency bias} commonly observed in human marking, particularly in borderline cases where partial credit decisions typically favor students\cite{tomic2025anxiety}. The AI system's inability to consider contextual factors such as student effort, improvement over time, or exceptional circumstances results in more rigid rubric application.

\subsection{Distributional Similarity and Educational Implications}
Despite scoring differences, both assessment approaches demonstrate similar left-skewed distributions (human: -1.724, AI: -1.409), indicating shared recognition of submission quality hierarchies. This distributional alignment suggests that while the systems may disagree on absolute scores, they generally distinguish between higher and lower quality work in consistent patterns.

The broader distribution of AI scores may actually provide pedagogical advantages by creating greater score differentiation among submissions. Human scores concentrate heavily in the 80-100 range, potentially limiting the ability to distinguish between different levels of competency. The AI system's wider score distribution could offer more granular feedback on relative performance, helping students better understand their position within the cohort.

\subsection{Feedback Quality and Specificity}
The AI system's primary strength lies in feedback generation rather than scoring accuracy. Table~\ref{tab:feedback_comparison} demonstrates marked differences in feedback specificity and technical detail. While human feedback often provides generic comments (\textit{"Good documentation and coding skills"}), AI-generated feedback identify specific technical implementations and explain why particular approaches succeed or fail.

\begin{table}[h!]
\centering
\caption{Comparison of Human and AI-Generated Feedback} \label{tab:feedback_comparison}
\begin{tabular}{p{3.3cm} p{4.4cm}}
\toprule
\textbf{Human Feedback} & \textbf{AI-Generated Feedback} \\
\midrule

Good documentation and all elements of marking rubric addressed. Excellent re-use of code and documentation. & The evaluation cannot be completed as the code and markdown content for both problems are missing. Without access to the actual code and markdown, it is impossible to assess the quality of function docstrings, variable naming conventions, code structure, readability, or markdown explanations. \\
\midrule

Good documentation, coding skills and use of function. Good to see attention to rubrics and the analysis is fine.  & The student's implementation of the optimize\_gas\_injection function is well-structured and meets most of the requirements. The student uses the SLSQP method from scipy.optimize.minimize, which is suitable for constrained optimization problems. \\
\midrule

Comments around appropriate choice of the model would have helped. Use of penalty is not required and you should have used ineq/eqs. Zero gas injection and yet oil extraction. & The student's analysis provides a comprehensive interpretation of the fitted parameters in the context of a logistic growth model. They correctly identify and explain the physical meaning of each parameter: K as the carrying capacity, r as the growth rate per hour, and A0 as the initial optical density. \\
\midrule

Provide documentation in terms of what the next function does. Zero gas resulting in oil extraction. Please look into the results always. Why would you inject gas if it is not resulting in oil extraction? & The student's implementation of the calculate\_gradients function meets all the specified criteria. The function exists with correct signature and the gradients for parameters a, c, and d are correctly calculated using the provided formulas. \\
\midrule

Excellent understanding of the process with close attention to marking rubrics. Plots could have been improved for better visibility. & The code demonstrates a consistent effort in providing documentation and maintaining code quality. All functions have docstrings, which is commendable. However, the format is inconsistent, and some descriptions are vague or repetitive. \\
\midrule

No explanation on how ChatGPT was used, but thorough reference list was provided. The residual plot could have been explained more. & Most functions have basic comments explaining their purpose, but they lack detailed docstrings that describe parameters, return values, and exceptions. This makes it harder for someone unfamiliar with the code to understand the full functionality. \\
\midrule

Minor gaps in explaining mathematical reasoning behind normalization approach. Some comments could be more detailed about return values and parameter types. & Most functions have docstrings that describe their purpose and parameters. However, some docstrings could be more detailed, especially in explaining the return values and the context of use. \\
% \midrule

% The dataset does not make sense. The model does not converge. No formulas are provided for the gradient computation. & The function calculate\_gradients is defined with the correct signature and the gradients for parameters a, c, and d are correctly calculated using the provided formulas. The implementation meets all specified criteria. \\
\bottomrule
\end{tabular}
\end{table}

\section{Limitations and Future work}
Several limitations constrain interpretation of these results. 

The sample size  disparity ($n=184$ human, $n=79$ AI) reflects the pilot deployment scope and  prevents direct statistical comparison. Additionally, the conservative AI 
scoring pattern may reflect overly stringent prompt calibration rather than 
inherent assessment accuracy, suggesting need for prompt optimization in 
future iterations.

Three key extensions will enhance the platform's impact: (1) Multi-language support to accommodate Java, C++, and other programming languages through language-specific testing frameworks and adapted LLM prompts; (2) Configurable LLM backends enabling institutions to select between local models (Qwen3, Kimi-K2-Instruct) and cloud APIs (GPT-4, Claude) based on performance and privacy requirements; and (3) Deep LMS integration with automatic bidirectional synchronization for Moodle, Blackboard, and Canvas to streamline instructor workflows and improve student experience.

\section{Conclusion}
The modular architecture maintains rubric alignment while generating detailed explanatory feedback, as evidenced by pilot deployment. While, AI scoring exhibits conservative tendencies compared to human marking, the system delivers superior feedback specificity that enables actionable student improvement. 

-----------------------------------------------------------------------
\bibliographystyle{IEEEtran}

\end{document}